\pdfoutput=1
\documentclass[conference]{IEEEtran}
\IEEEoverridecommandlockouts
\usepackage{cite}
\usepackage{csquotes}
\usepackage{caption}
\usepackage{subcaption}
\usepackage{amsmath,amssymb,amsfonts}
\usepackage{algorithmic}
\usepackage{graphicx}
\usepackage{textcomp}
\usepackage[svgnames, table, pdftex, dvipsnames]{xcolor}
\usepackage{xspace}
\usepackage{hyperref}
\usepackage{flushend}
\usepackage{tikz}
\usetikzlibrary{svg.path}

\newcommand\copyrighttext{%
  \footnotesize \textcopyright~2022 IEEE. Personal use of this material is permitted.
  Permission from IEEE must be obtained for all other uses, in any current or future
  media, including reprinting/republishing this material for advertising or promotional
  purposes, creating new collective works, for resale or redistribution to servers or
  lists, or reuse of any copyrighted component of this work in other works.
  }
\newcommand\copyrightnotice{%
\begin{tikzpicture}[remember picture,overlay]
\node[anchor=south,yshift=20pt] at (current page.south) {\fbox{\parbox{\dimexpr\textwidth-\fboxsep-\fboxrule\relax}{\copyrighttext}}};
\end{tikzpicture}%
}

\begin{document}

\title{How to Evaluate Explainability? -- \\ A Case for Three Criteria
}

\author{\IEEEauthorblockN{
Timo Speith}
\IEEEauthorblockA{Saarland University, Institute of Philosophy, Saarbrücken, Germany}
\IEEEauthorblockA{Saarland University, Department of Computer Science, Saarbrücken, Germany}

Email: timo.speith@uni-saarland.de
}

\maketitle

\copyrightnotice
\vspace{-2ex}

\begin{abstract}
The increasing complexity of software systems and the influence of software-supported decisions in our society have sparked the need for software that is safe, reliable, and fair. Explainability has been identified as a means to achieve these qualities. It is recognized as an emerging non-functional requirement (NFR) that has a significant impact on system quality. However, in order to develop explainable systems, we need to understand \emph{when} a system satisfies this NFR. To this end, appropriate evaluation methods are required. However, the field is crowded with evaluation methods, and there is no consensus on which are the \enquote{right} ones. Much less, there is not even agreement on which criteria should be evaluated. In this vision paper, we will provide a multidisciplinary motivation for three such quality criteria concerning the information that systems should provide: \emph{comprehensibility}, \emph{fidelity}, and \emph{assessability}. Our aim is to to fuel the discussion regarding these criteria, such that adequate evaluation methods for them will be conceived. 

\end{abstract}

\begin{IEEEkeywords}
Explainability, Explainable Artificial Intelligence, Evaluation, Non-Functional Requirements, NFR, XAI
\end{IEEEkeywords}

\section{Introduction}

In today's world, software systems are increasingly impacting sensitive areas of daily life (e.g., healthcare \cite{Holzinger2019Causability} or criminal justice \cite{Dressel2018Accuracy}). However, contemporary systems are complex, effectively being black boxes for the growing number of stakeholders involved. The ubiquitous influence of such \enquote{black-box systems} has induced discussions about the transparency and ethics of modern systems \cite{Adadi2018Peeking}. Responsible collection and use of data, privacy, safety, and security are just a few among many concerns. In light of this, it is becoming increasingly crucial to understand how to incorporate these concerns into systems and, thus, how to deal with them during software engineering (SE) and requirements engineering (RE) \cite{Chazette2021Exploring}.

\subsection{The Need for Explainability}

In this regard, research regarding \emph{explainability} has recently experienced an upsurge \cite{Arrieta2020Explainable, Langer2021What}. Explainability promises to alleviate a system's lack of transparency \cite{Chazette2020Explainability} and to provide a fruitful way to address ethical concerns about modern systems \cite{Langer2021Auditing}. Furthermore, explainability is increasingly seen as crucial for high software quality, and should be treated as a non-functional requirement (NFR) \cite{Koehl2019Explainability}. Unfortunately, \enquote{explainability} is a nebulous and elusive concept that is hard to target. This causes difficulties for researchers, especially when it comes to designing and evaluating explainability in systems  \cite{Langer2021What, Krishnan2020Against}. Regarding the \emph{design} of explainable systems, for instance, scholars often seem to rely on intuition \cite{Chazette2021Exploring, Miller2017Explainable}.

As for the \emph{evaluation} of a system's explainability, there is no consensus on what constitutes a good method \cite{Vilone2021Notions, Zhou2021Evaluating, Brunotte2022Explainability}. There are several methods for evaluating explainability, each of which comes with its own underlying rationale for which criteria are essential for good system explainability. However, these evaluation methods all have their limitations. First and foremost, we are not aware of any theoretical considerations that motivate certain quality criteria over others.

In this vision paper, we will provide a multidisciplinary motivation for three quality criteria concerning the information that systems should provide in order to count as explainable: \emph{comprehensibility}, \emph{fidelity}, and \emph{assessability}. Our aim is to to fuel the discussion regarding these criteria, such that useful and adequate evaluation methods for them will be conceived.

\subsection{The Goals of Explainability}

\begin{figure}[h]
\vspace{-1ex}
\centering
\begin{tikzpicture}[scale=0.55, transform shape,
    block1/.style={rectangle, draw=black, thick, text width=8em, align=center, rounded corners, minimum height=3em},
    block2/.style={rectangle, draw=black, thick, text width=6em, align=center, rounded corners, minimum height=3em},]
    
    \node[draw, block2] at (-5, 0)   (xai) {Explainability Approach};
    \node[draw, block2] at (0, 0)    (exp) {Explanatory Information};
    \node[draw, block2] at (2.45, 2) (und) {Understanding};
    \node[draw, block2] at (5, 0)    (des) {Desiderata Satisfaction};
    
    \draw[-latex] (xai) -- (exp)  node[midway, below] (fac) {provides};
    
    \draw[-latex] (exp) -- (und)  node[midway, left=1ex] (let) {facilitates};
    \draw[-latex] (und) -- (des)  node[midway, right=1ex] (sat) {affects};
    \draw[-latex, dotted] (exp) -- (des)  node[midway, below] (med) {};
    
    
\end{tikzpicture}
\caption{A simplified version of the models of the main concepts and relations in explainability that Langer et al. \cite{Langer2021What} and Hofmann et al. \cite{Hoffman2018Metrics} propose. Dotted arrows indicate relations that are fully mediated by the solid arrows.}
\label{fig:model}
\vspace{-1ex}
\end{figure}
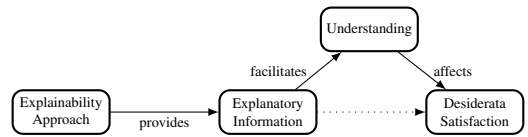

Before we can motivate that these criteria must be evaluated, we first need to be clear about the goals of explainability in general. Hofmann et al. \cite{Hoffman2018Metrics} and Langer et al. \cite{Langer2021What} propose similar models that outline how key concepts in explainability are related to one another. According to their models, \emph{explainability approaches} (i.e., ways to reach explainability) provide \emph{explanatory information} with the aim of facilitating people's \emph{understanding}. This understanding, in turn, affects the satisfaction of so-called \emph{desiderata} (e.g., transparency, safety; see Figure \ref{fig:model} for a simplified version of their models). 

To summarize the models, explainability does not necessarily serve as a goal directly, but rather as a means to other goals (i.e., desiderata), such as transparency, satisfaction, and safety. In a literature survey of over 100 works, Langer et al. \cite{Langer2021What} found 29 desiderata that explainability is intended to contribute to. In an even more comprehensive literature review of over 200 works, Chazette et al. \cite{Chazette2021Exploring} found 57 such desiderata 
that can (positively \emph{and} negatively) be impacted by explainability and should be considered when it comes to high system quality.

\section{Quality Criteria for Evaluating Explanatory Information}
\label{sec:critexpl}

Based on the above conceptualization of the goal(s) of explainability, it is possible to distill multidisciplinary motivated quality criteria that the explanatory information provided by systems should meet to ensure good system explainability.

\subsection{The Three Dimensions of Explainability}

In particular, we motivate three criteria by which to evaluate the quality of explanatory information produced by an explainability approach. These criteria are inspired by the \emph{three dimensions of explainability} as described by Baum et al. \cite{Baum2017Challenges}:

\begin{itemize}
\item \textbf{Comprehensibility:} Explanatory information must be conveyed in a way that is comprehensible to humans, for humans must understand (certain aspects of) the system based on the information. Obviously, comprehensible information is best suited for this purpose.

\item \textbf{Fidelity:} Explanatory information must be fidelitous with respect to (the aspect of) the system it is about. For instance, the information must describe the accurate (i.e., correct, real) reasons for a system acting the way it did. Although some details may be omitted or simplified for comprehensibility, outright lies should never be told.

\item \textbf{Assessability:}\footnote{In the original source, the third dimension is \emph{Permissibility}. We deliberately deviate here, as we are concerned with the desirable \emph{properties of explanatory information}. The original source dealt with the \emph{properties of computational systems} that are necessary to make them trustworthy. Trustworthy systems should indeed not only be be able to deliver assessable information but they should in fact be permissible. That is, they should be positively assessed concerning certain desiderata (e.g., morality, fairness, safety, reliability). However, the information must be assessable with respect to these desiderata to allow somebody to determine whether the system is in fact permissible.} Explanatory information must be such that one can assess the satisfaction of a given desideratum. In other words, the quality of explainability depends on the desiderata that one aims to satisfy with it.
\end{itemize}

While these criteria are inspired by previous work (i.e., \cite{Baum2017Challenges}), it is easy to motivate them independently. First, information that is incomprehensible or misleading (because it is not fidelitous) is unlikely to help facilitate a person's true understanding of a phenomenon. Considering the fact that understanding is an essential factor in the field of explainability \cite{Langer2021What, Chazette2021Exploring, Koehl2019Explainability}, systems that produce incomprehensible or infidelitous information do not count as explainable.

Let us now turn to assessability. As stated above, the final goal of receiving information about (certain aspects of) a system is to satisfy certain desiderata \cite{Langer2021What, Chazette2021Exploring, Arrieta2020Explainable}. However, it is conceptually possible that a given piece of information can be both comprehensible and fidelitous, but completely irrelevant to the desiderata of interest. For this reason, we need a criterion that restricts the set of relevant pieces of information to those that serve the the targeted goal: satisfying these desiderata.

\subsection{Arguments for Our Criteria}

In what follows, we will provide multidisciplinary arguments to further corroborate our criteria. In particular, these arguments are based on satisfying certain often cited desiderata (see \cite{Langer2021What, Langer2021Auditing, Chazette2021Exploring}) that are important for different disciplines.

\subsubsection{Acceptance and Trustworthiness}

Our primary argument is the need for \emph{acceptance} and \emph{trustworthiness}. It is often argued that systems that are unable to explain themselves will lack \emph{acceptance} in the long run \cite{Bilgic2005Explaining, Krishnan2020Against, Pieters2011Explanation}.

From a \emph{moral} point of view, it is plausibly to assume that the deployment of some kinds of (autonomous) systems promises to bring about overall positive effects. Thus, as long as people do not accept these systems, their presumably beneficial deployment is threatened \cite{Baum2018From, Baum2018Towards}.

Furthermore, from a \emph{pragmatical} point of view, developers and companies want the systems they design to be accepted by the populace. Systems that are developed but not sufficiently deployed will threaten the economical basis of the company (and, thus, of the developers in it).


For people to accept systems, \emph{trust} is an essential prerequisite \cite{Rosenfeld2019Explainability, Glass2008Toward}. However, not any kind of trust will be beneficial in the long run. Only \emph{justified} trust in a system will bring about the best consequences concerning it; inadequate trust can lead to disastrous consequences \cite{Kaestner2021Relation}. Consequently, adequate trust is desirable from a moral and pragmatical point of view.

So, to significantly increase the probability that autonomous systems will be used on a large scale, they must be justifiably trusted. Justified trust is based on trustworthiness \cite{Kaestner2021Relation, Jacovi2021Formalizing}. Furthermore, trustworthiness depends on a stakeholder's justification in believing that a system works properly, and explanations are one way in giving these justifications \cite{Kaestner2021Relation}. Thus, explanations are an important factor in calibrating trust, a connection that is is often argued for \cite{Bilgic2005Explaining, Herlocker2000Explaining, Sinha2002Role, Symeonidis2009MoviExplain}.

Of course, not just any kind of explanation will do. To be trustworthy, software systems must be able to justify their actions \emph{in the right way}. A well-motivated means of doing so is to provide information that at least meets the criteria above.

First, the information must be comprehensible, because incomprehensible information is not sufficient to provide any justification. Furthermore, the information must be fidelitous, since a lying system, even if otherwise functioning properly, cannot count as trustworthy. Finally, the information must support the assessability of pertinent desiderata, as this allows people to judge whether the system is working properly.\footnote{For further discussion of why the quality criteria we set out are good criteria to judge the trustworthiness of a system, see \cite{Baum2017Challenges}.}

\bigskip
\subsubsection{Accountability and Autonomy}

Our second argument is the need for \emph{responsibility} and \emph{autonomy} in the interaction of artificial systems and humans. It is foreseeable that in the future, humans will increasingly rely on the decisions of software systems. Currently, the trend is to use such systems to make recommendations in situations where much is at stake.


From a \emph{legal} and \emph{moral} perspective, it is highly desirable to be able to hold someone accountable when something goes wrong in such critical situations \cite{Hagendorff2020Ethics, Wachter2017Transparent}. However, with modern systems, it is often difficult to assign responsibility \cite{Baum2022Responsibility, Matthias2004Responsibility}. In order to still be able to hold someone accountable in such situations, it is often assumed that there must be a human in the loop who makes the final decision and who is the most probable bearer of responsibility \cite{Baum2022Responsibility}.

Imagine an HR person. They receive a recommendation for each and every applicant, solely stating whether to keep the applicant in the running or not. How can the HR person come to the kind of decision that makes them a bearer of responsibility? If they decide solely in line with the recommendations, they are just a submissive executor of the decision made by a system. In this case, one could simply dismiss the human in the loop altogether. If they decide against the recommendation, they cannot have good reasons for doing so without knowing the reasons for the recommendation in the first place \cite{Baum2022Responsibility}.

Accordingly, to properly bear responsibility for a decision, it must be made \emph{autonomously} \cite{DeLaat2018Algorithmic}. However, without being able to \emph{comprehend} and \emph{assess} a recommendation, human beings lose their autonomy. After all, when someone is to decide competently and autonomously, they need more than simple recommendations from a system. A human in the loop needs reasons for the recommendations to assess their correctness and potentially challenge them. Explanatory information, thus, must be at least \emph{comprehensible} and \emph{assessable} to properly allow for human autonomy, and, with this, human accountability.


\subsubsection{Fairness}

The last argument is the need for \emph{fairness}. In many situations, human beings are immediately affected by the decisions made or supported by software systems \cite{Langer2021What, Langer2021Auditing}.

From a \emph{societal} point of view, a purely statistical checking of systems is not sufficient to ensure that no individual's fundamental rights are violated. Instead, one needs to be able to assess the \emph{individual} decisions of such systems. In other words, one should not only be concerned with whether a system overall did not discriminate against certain groups of people, but rather with whether each decision did not do so.

The reason for this restriction is simple: different systems can arrive at the same result in completely dissimilar ways. To illustrate, imagine two systems that rank job applicants. Both systems rank a Black woman last, but for different reasons. The first system does so because she really is the least qualified (e.g., she has no prior work experience, bad grades, etc.). The second system does so just because she is Black and a woman. 

While to some this point may seem to be practically unimportant under the assumption that a statistical bias can be reasonably excluded, it is essential to make the system trustworthy and establish public acceptance. In particular, it is crucial to guarantee that the fundamental values of liberal societies are considered sufficiently. Statistical adherence to norms and rules is important, but it should not be the only thing to be considered. Each individual case matters. Explanatory information, thus, must be at least \emph{assessable} and \emph{fidelitous}.

\section{Evaluating Contemporary Approaches}

Having motivated and argued for our criteria, we will now turn to their application. In particular, we will examine some prominent explainability approaches in the field of explainable artificial intelligence (XAI) and evaluate the information they produce based on our criteria. In this regard, it should be noted that we are mostly raising theoretical points, since, as stated above, there is little to no agreement on evaluation methods. Still, this should show that our criteria are worthy of attention.

\begin{figure}[htbp]
\vspace*{-2.5ex}
\centering
\begin{subfigure}{.45\linewidth}
    \centering
    \includegraphics[width=\linewidth]{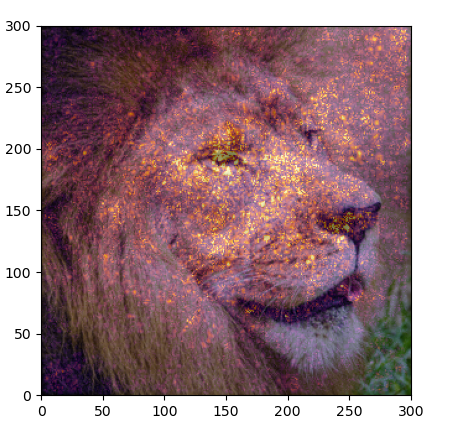}
    \vspace{-3.25ex}
    \caption{Vanilla Backpropagation}
    \label{fig:vanilla}
\end{subfigure}%
\begin{subfigure}{.45\linewidth}
    \centering
    \includegraphics[width=\linewidth]{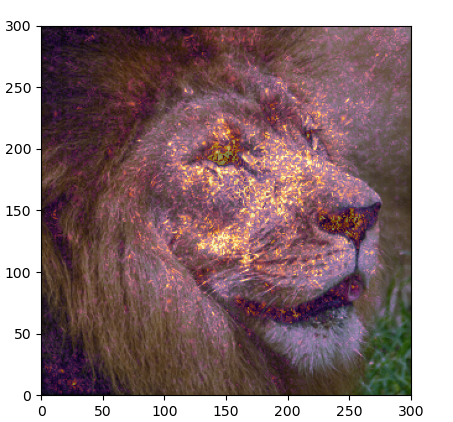}
    \vspace{-3.25ex}
    \caption{Integrated Gradients}
    \label{fig:integrated}
\end{subfigure}
\begin{subfigure}{.45\linewidth}
    \centering
    \includegraphics[width=\linewidth]{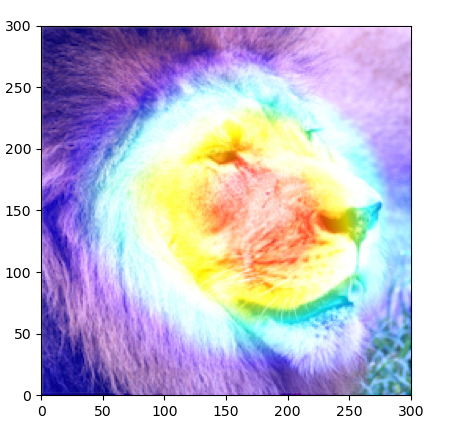}
    \vspace{-3.25ex}
    \caption{Grad-CAM}
    \label{fig:grad}
\end{subfigure}%
\begin{subfigure}{.45\linewidth}
    \centering
    \includegraphics[width=\linewidth]{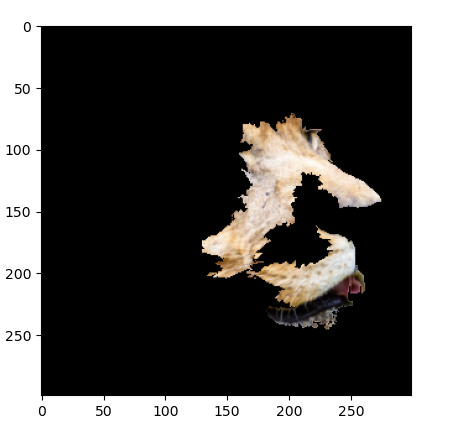}
    \vspace{-3.25ex}
    \caption{LIME}
    \label{fig:lime}
\end{subfigure}%
\vspace{-0.5ex}
\caption{Outputs of different explainability approaches.}
\vspace{-2ex}
\label{fig:methods}
\end{figure}

In particular, we will examine LIME \cite{Ribeiro2016Why}, Vanilla Backpropagation \cite{Simonyan2013}, Guided Backpropagation \cite{Springenberg2015Striving}, Integrated Gradients \cite{Sundararajan2017Axiomatic}, Grad-CAM \cite{Selvaraju2017Grad-CAM}, and TCAV \cite{Kim2018Interpretability}. 
Figure \ref{fig:methods} shows exemplary outputs of some of these approaches.

\subsection{Comprehensibility}
Most of the above approaches benefit from the visual representation of the explanatory information they produce. One does not have to be an expert to recognize that highlighted areas of an image correspond to important parts. However, people without sufficient background in machine learning (ML) will, in many cases, not comprehend what the highlighting means on a deeper (e.g., technical) level \cite{Gilpin2018Society, Langer2021What}.

In particular, pixel-oriented saliency masks (produced by, for instance, Vanilla Backpropagation and Integrated Gradients) suffer from this. Additionally, these techniques have further problems. First, they generate explanatory information that is very difficult to analyze even for experts \cite{Langer2021What, Gilpin2018Society, Alqaraawi2020Evaluating}. For instance, for people without sufficient background, Vanilla Backpropagation sometimes seems to produce saliency masks with random highlighting (see Figure \ref{fig:vanilla}). Moreover, these techniques suffer from computational artifacts, often producing out-of-context highlighting that hinders comprehensibility.

LIME and TCAV perform better in terms of comprehensibility. When it comes to LIME, its inventors have confirmed the comprehensibility of the information it generates in several studies \cite{Ribeiro2016Why}. As for TCAV, it has the advantage that the users themselves can define the concepts for which an artificial neural network (ANN) is to be tested. On the other hand, TCAV has the disadvantage that it requires at least some background knowledge of how ANNs work, since one must specify which layers of an ANN to test and analyze the results accordingly. For example, one needs to know that deeper layers of ANNs are more like to respond to complex concepts (e.g., gender) than other layers that respond to simpler concepts (e.g., color).

\subsection{Fidelity}
Currently, there is not much research that evaluates the fidelity of explanatory information produced by explainability approaches. The reason for this shortcoming is, among others, that it is not yet entirely clear, formally and technically, what it means for explanatory information to be fidelitous \cite{Molnar2019Interpretable, Adebayo2020Sanity, Amparore2021Trust}. 

Despite this difficulty, there are some authors who try to address the issue. For instance, Adebayo et al. \cite{Adebayo2020Sanity} developed a sanity check for approaches producing saliency maps. Using this test, they evaluated several well-known approaches. They found that while some approaches pass their sanity check and produce fidelitous information (e.g., Vanilla Backpropagation, Grad-CAM), others do not (e.g., Guided Backpropagation).

Another notable work is that of Amparore et al. \cite{Amparore2021Trust}. Among others, they found that many implementations of LIME do not satisfy the theoretical properties originally promised by this approach. For instance, these implementations produce unstable and infidelitous information. The infidelity of LIME was to be expected, since it is a \emph{model-agnostic} approach. In other words, the internals of the model to be explained are not taken into account in the generated information: it is based solely on an input-output analysis \cite{Speith2022Review}.

\subsection{Assessability}
It should be noted that assessability is a difficult criterion to test because there are so many desiderata that could, in principle, be of interest. For this reason, we will only give some general remarks, starting with LIME.

In addition to studies on the comprehensibility of the information generated by LIME, Ribeiro et al.  \cite{Ribeiro2016Why} have also conducted studies on what participants could do with this information. Among others, these studies showed that using LIME enables individuals to identify incorrect classifications and even helps them improve the classifier. This suggests that the provided information is useful for assessing at least some desiderata (e.g., correctness, trustworthiness, and fairness).

Kim et al. \cite{Kim2018Interpretability} compare different approaches that generate saliency maps (i.e., Vanilla Backpropagation, Guided Backpropagation, Integrated Gradients, and Smoothgrad) to check whether the information they generate enables one to evaluate whether a given classification makes sense.

To this end, they add a visible label, which varies across trials, to the lower left corner of images. In one trial, the label is constant across individual classes (e.g., all images of cabs receive the label \enquote{cab}). In another trial, the label sometimes deviates within a class (e.g., some cabs receive the label \enquote{cucumber}). In yet another trial, the label is completely random (e.g., each image of a cab receives a different label). They found that in all trials, the image's lower left corner was highlighted by the approaches to a non-negligible extent.

It is important to point out that this does not preclude the approaches from being fidelitous to the classification algorithm, since the lower left corner might actually be used by it (it has a prominent label, after all), even if only insignificantly. However, it does mean that it is difficult to impossible to use the generated information for meaningful assessments.

Coming to the next approaches, Rudin \cite{Rudin2019Stop} argues that many heatmaps (including those produced by Grad-CAM) do not really allow for good assessments. The main reason for this is that the heatmap for the most probable class (e.g., dog) is often hardly distinguishable from that of a less probable class (e.g., wolf). Accordingly, it is questionable whether the obtained information allows for assessing certain desiderata.

Finally, let us talk about TCAV. In our opinion, TCAV performs best when it comes to assessability. This is the case because it allows for hand-crafted concepts to be reviewed. In other words, a person using TCAV is not simply confronted with an unchanging set of information, as with other approaches, but can inquire after the information that is of interest (e.g., whether gender played a role in classification).

\renewcommand{\arraystretch}{1.2}
\rowcolors{2}{blue!20}{white}
\begin{table}[htbp]
\vspace{-0.75ex}
    \centering
    \begin{tabular}{|c|c|c|c|}
        \hline
        \rowcolor{blue!50}
        Approach & Comprehens. & Fidelity & Assessability \\
        \hline
        LIME & $+$ & $--$ & $+$ \\
        Vanilla Backpropagation & $--$ & $+$ & $-$ \\
        Guided Backpropagation & $-$ & $-$ & $-$ \\
        Integrated Gradients & $-$ & ? & $-$ \\
        Grad-CAM & $+$ & $+$ & $-$ \\
        TCAV & $+$ & $+$ & $++$\\
        \hline
    \end{tabular}
    \smallskip
    \caption{Contemporary explainability approaches evaluated.}
    $+$ indicates a positive evaluation of this criterion, $-$ a negative one.
    \vspace{-0.75ex}
    \label{tab:evaluation}
\end{table}

Summarizing the above, Table \ref{tab:evaluation} offers an overview of how we believe the explanatory information generated by the discussed approaches fare in terms of our criteria. Overall, we think that TCAV fares best, as it is comprehensible, fidelitous, and allows for a wide range of assessments.

\section{Conclusion and Future Work}

In this vision paper, we have motivated and argued for three quality criteria of the information that systems should provide in order to be considered explainable: \emph{comprehensibility}, \emph{fidelity}, and \emph{assessability}. By  applying these criteria to the outputs of some well-known approaches in the field of XAI, we have substantiated the reasonableness of these criteria.

While there are some evaluation methods for the comprehensibility of explanatory information (e.g., some items on the \emph{explanation satisfaction scale} by Hoffman et al. \cite{Hoffman2018Metrics}), there is much less literature on evaluation methods for fidelity and assessability. Thus, if one accepts our quality criteria, there is much to be done about evaluation methods.

Among others, one should become clear about what \emph{fidelity} actually means and formalize this. Moreover, it should be further specified what it means for information to be \emph{assessable}. In general, we hope for more research on evaluation methods.

Finally, these are only three criteria among many others that are reasonable (e.g., conciseness). By giving arguments for these criteria, we have taken a principled approach and given them a theoretical motivation and foundation. We hope that more criteria will be underpinned in this way in the future.

\section*{Acknowledgments}
Work on this paper was funded by the Volkswagen Foundation grant \textsc{AZ 98514} \href{https://explainable-intelligent.systems}{\enquote{Explainable Intelligent Systems}} (\textsc{EIS}) and by the \textsc{DFG} grant 389792660 as part of \href{https://perspicuous-computing.science}{\textsc{TRR}~248}. 

\bibliographystyle{IEEEtran}
\bibliography{bibliography}

\end{document}